\documentclass[journal]{IEEEtran}
 \usepackage{amsmath}
 \usepackage{subfigure}
 \usepackage{graphicx,graphics,color,psfrag}
 \usepackage{cite,balance}
 \usepackage{amsmath,amssymb,amsfonts}
 \usepackage{caption}
 \usepackage{amsmath}
 \usepackage{amsthm}
\usepackage{textcomp}
 \allowdisplaybreaks
 \usepackage{algorithm}
  \usepackage{algorithmic}
 \usepackage{accents}
 \usepackage{amsthm}
 \usepackage{bm}
  \usepackage{url}
 \usepackage[english]{babel}
 \usepackage{multirow}
 \usepackage{enumerate}
 \usepackage{cases}
 \usepackage{stfloats}
 \usepackage{dsfont}
 \usepackage{color,soul}
 \usepackage{amsfonts}
 \usepackage{cite,graphicx}
 \usepackage{subfigure}
 \usepackage{fancyhdr}
 \usepackage{hhline}
 \usepackage{graphicx,graphics}
 \usepackage{array,color}
  \usepackage{bm}

\newtheorem{lemma}{\emph{\underline{Lemma}}}

\def\l{\left}
\def\r{\right}
\def\({\left(}
\def\){\right)}

\def\b0{{\mathbf{0}}}

\newcommand{\tr}{\mathrm{tr}}
\newcommand{\diag}{\mathrm{diag}}

\newcommand{\nn}{\nonumber}

\begin{document}
\captionsetup[figure]{name={Fig.}}

\title{\huge 
Wireless Communication via Double IRS: Channel Estimation and Passive Beamforming Designs} 
\author{{Changsheng You,~\emph{Member,~IEEE}, Beixiong Zheng,~\emph{Member,~IEEE},
	 and Rui Zhang,~\emph{Fellow,~IEEE}} 
	 	 \vspace{-14pt}
		   \thanks{\noindent The authors are with the Department of Electrical and Computer Engineering, National University of Singapore, Singapore 117583 (Email: \{eleyouc, elezbe, elezhang\}@nus.edu.sg).\vspace{-12pt}
}}
\maketitle

\begin{abstract}
In this letter, we study efficient channel estimation and passive beamforming designs for a double-intelligent reflecting surface (IRS) aided single-user communication system, where a user communicates with an access point (AP) via the cascaded user-IRS $1$-IRS $2$-AP \emph{double-reflection} link. 
First, a general channel estimation scheme is proposed for the system under any arbitrary inter-IRS channel, where all coefficients of the cascaded channel are estimated. 
  Next, for the typical scenario with a line-of-sight (LoS)-dominant inter-IRS channel, we propose another customized scheme to estimate two signature vectors of the rank-one cascaded channel with significantly less channel training time than the first scheme. For the  two proposed channel estimation schemes, we further optimize their corresponding cooperative  passive beamforming for data transmission to maximize the achievable rate with the training overhead and channel estimation error taken into account. Numerical results show that deploying two cooperative IRSs with the proposed channel estimation and passive beamforming designs achieves significant rate enhancement as compared to the conventional case of single IRS deployment.
 \vspace{-4pt}
 \end{abstract}
\begin{IEEEkeywords}
Intelligent reflecting surface, cooperative passive beamforming, channel estimation.
\vspace{-5pt}
\end{IEEEkeywords}

\vspace{-5pt}
\section{Introduction}
Intelligent reflecting surface (IRS) has emerged as a promising technology to enhance the spectral efficiency of wireless communication systems cost-effectively, by smartly controlling signal reflection via a massive number of low-cost passive reflecting elements \cite{wu2020intelligent,basar2019wireless}.
 Moreover, IRS is generally of low weight and energy consumption, thus can be easily coated on environmental objects to ubiquitously engineer the radio propagation environment effectively. 

The exiting works on IRS have mostly considered the wireless communication systems aided by one single IRS  or multiple distributed  IRSs (e.g., \cite{wu2019intelligent,yu2020robust,ning2019channel}), each independently assisting the communication of its surrounding users with their associated access points (APs). To reap the IRS passive beamforming gain, different IRS channel estimation schemes have been proposed in the literature to acquire the channel state information (CSI) of the cascaded user-IRS-AP single-reflection link \cite{OFDM_BX,you2019progressive,you2020fast}.  However, these 
designs are inapplicable to  the practical scenario where the signal reflected by one single IRS cannot bypass all the main obstructions, e.g.,
 the communication along corridor corners as illustrated in Fig.~\ref{Fig:Syst}. In this case, two or more IRSs need to form a cooperative network to establish a blockage-free link from transmitter to receiver via multiple signal reflections.  
Although  an initial attempt has been made in \cite{han2020cooperative} to address this issue by jointly  designing the passive beamforming for a double-IRS aided communication system, it assumed the line-of-sight (LoS) channel model for all involved links and perfect CSI available at the AP. This, however, 
simplified two challenging issues for implementing the cooperative double-IRS system in practice, namely, its cascaded channel estimation and optimal passive beamforming with the channel estimation error taken into account under the general Rician fading channel model.



To address the above issues, we consider in this letter a double-IRS cooperatively aided communication system as illustrated in Fig.~\ref{Fig:Syst}, where a user can communicate with the AP through the double-reflection (i.e., user-IRS $1$-IRS $2$-AP) link only. We first consider the general scenario with any arbitrary inter-IRS (i.e., IRS $1$-IRS $2$) channel and propose an efficient scheme to estimate all coefficients of the cascaded channel. Next, for the typical scenario with an LoS-dominant  inter-IRS channel  in practice, we propose another customized scheme to estimate two signature vectors of the rank-one cascaded channel with significantly less channel training time than the first scheme. For the two proposed channel estimation schemes, we further optimize their corresponding cooperative  passive beamforming for data transmission based on the estimated channel to maximize the achievable rate with the training overhead and channel estimation error taken into account. Numerical results show that deploying two cooperative IRSs with the proposed  channel estimation and passive beamforming designs achieves significant rate enhancement as compared to the conventional design of  placing all reflecting elements on one single IRS in the vicinity of the user that may suffer severe signal attenuation  in complex propagation environment.
\vspace{-12pt}
\section{System Model}\label{Sec:Model}
\begin{figure}[t]
\vspace{2pt}
\begin{center}
\includegraphics[height=3cm]{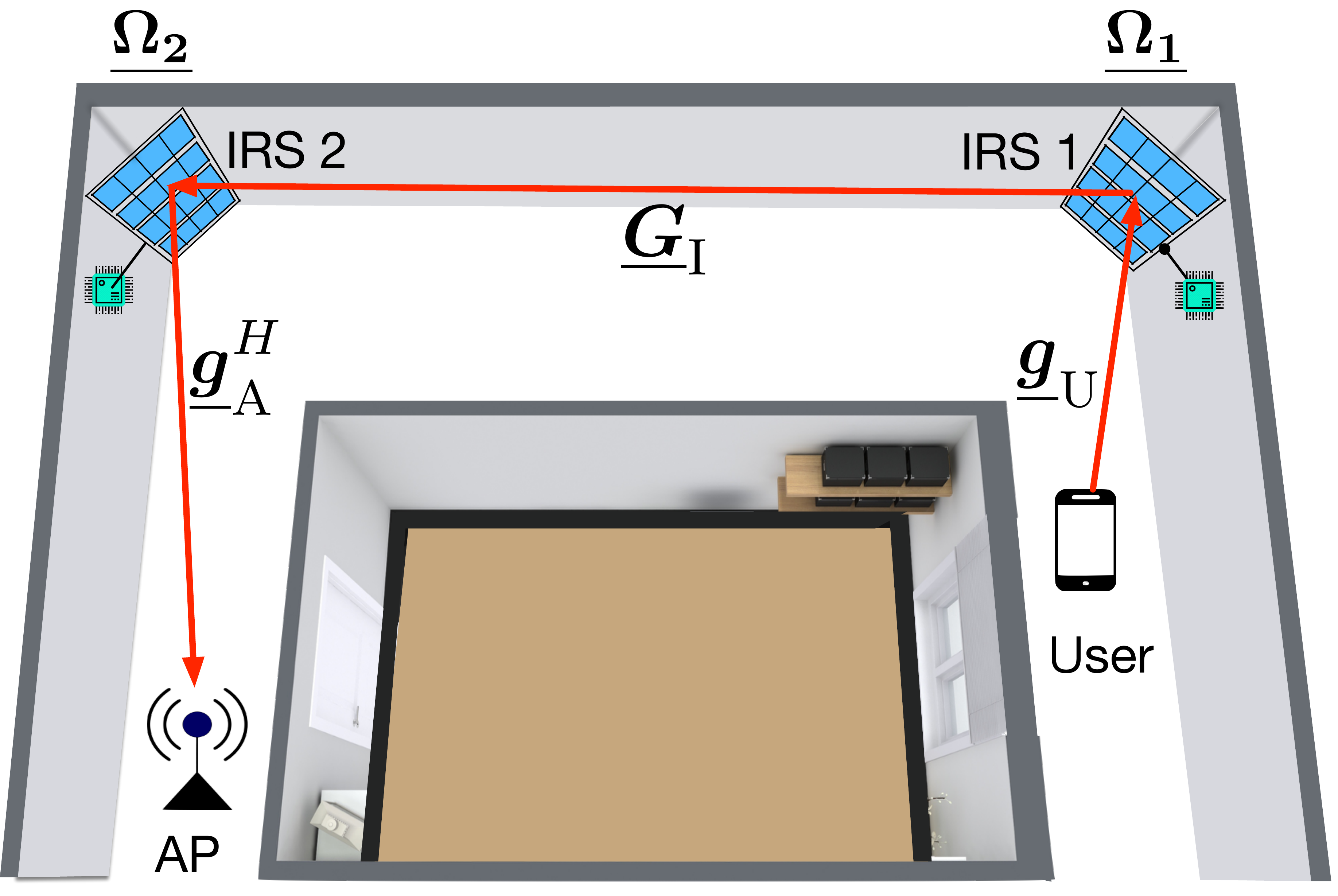}
\caption{A double-IRS cooperatively aided single-user communication system.}
\label{Fig:Syst}
\end{center}
\vspace{-7pt}
\end{figure}

Consider a double-IRS cooperatively aided single-user communication system shown in Fig.~\ref{Fig:Syst}, where two IRSs are properly deployed to assist in the uplink communication from a user to an AP, both  equipped with a single antenna. We focus on a challenging scenario  where the (direct) user-AP, IRS $1$-AP, and user-IRS $2$ links are blocked due to e.g., the corners along a corridor as illustrated in Fig.~\ref{Fig:Syst}; thus the user can be served  through the cascaded user-IRS $1$-IRS $2$-AP link only. To simplify the design complexity for the double-IRS  channel estimation and passive beamforming, we  group the $N_1$ ($N_2$) reflecting elements of IRS $1$ (IRS $2$)  into $M_1$ ($M_2$) sub-surfaces, each consisting of $N_0$ adjacent elements sharing a common reflection coefficient \cite{OFDM_BX}. 
Moreover, each IRS is attached with a smart controller for adjusting signal reflection at its reflecting elements as well as exchanging information with the  AP via a separate reliable backhaul link.

{\bf Channel model:}
We assume the quasi-static block fading model for all channels involved and focus on the uplink communication in one particular fading block.
To obtain the \emph{group-wise} channel model, we first denote $\underline{\boldsymbol{g}}_{\rm U}\in\mathbb{C}^{N_1\times1}$, $\underline{\boldsymbol{G}}_{\rm I}\in\mathbb{C}^{N_2\times N_1}$, and $\underline{\boldsymbol{g}}_{\rm A}^{H}\in\mathbb{C}^{1\times N_2}$ as the \emph{element-wise}  baseband channels from the user to IRS $1$, IRS $1$ to IRS $2$, and  IRS $2$ to the AP, respectively, all of which are assumed to follow the Rician fading. For example, the inter-IRS channel, $\underline{\boldsymbol{G}}_{\rm I}$, can be modeled by $\underline{\boldsymbol{G}}_{\rm I}=\sqrt{K_{\rm I}/(1+K_{\rm I})} \underline{\boldsymbol{G}}_{\rm I}^{\rm L}+\sqrt{1/(1+K_{\rm I})} \underline{\boldsymbol{G}}_{\rm I}^{\rm N}$,
where $K_{\rm I}$ is the Rician factor of $\underline{\boldsymbol{G}}_{\rm I}$, and $\underline{\boldsymbol{G}}_{\rm I}^{\rm L}$ and $\underline{\boldsymbol{G}}_{\rm I}^{\rm N}$ denote the deterministic LoS and random non-LoS (NLoS) Rayleigh fading components, respectively.
For each IRS $k\in\{1,2\}$, we denote by $\underline{\boldsymbol{\Omega}}_k\triangleq\diag(e^{\jmath\underline{\omega}_1},  \cdots,  e^{\jmath\underline{\omega}_{N_k}})\in\mathbb{C}^{N_k\times N_k}$ its diagonal reflection matrix, where we assume for simplicity that the reflection amplitude of each element is set to be one and $\underline{\omega}_n\in[0, 2\pi)$, $n\in\{1,\cdots, N_k\}$ denotes the phase shift of element $n$.
As such, the equivalent single-input-single-output (SISO) channel from the user to AP is given  by
\begin{align}
\vspace{-10pt}
g=  \underline{\boldsymbol{g}}_{\rm A}^{H} \underline{\boldsymbol{\Omega}}_2 \underline{\boldsymbol{G}}_{\rm I} \underline{\boldsymbol{\Omega}}_1 \underline{\boldsymbol{g}}_{\rm U}=\underline{{\boldsymbol \theta}}_2^{H}  \underline{\boldsymbol{H}}~ \underline{{\boldsymbol \theta}}_1,
\label{Eq:Effe2}
	 \vspace{-10pt}
\end{align}
where  $\underline{\boldsymbol{\theta}}_k\triangleq\l[e^{\jmath\omega_1}, \cdots, e^{\jmath\omega_{N_k}}\r]^T, \forall k\in\{1,2\}$; and $\underline{\boldsymbol{H}}\triangleq\diag(\underline{\boldsymbol{g}}_{\rm A}) \underline{\boldsymbol{G}}_{\rm I} \diag(\underline{\boldsymbol{g}}_{\rm U})\!\in\!\mathbb{C}^{N_2\times N_1}$
denotes the element-wise cascaded user-IRS $1$-IRS $2$-AP channel without IRS phase shifts. Based on the element-grouping strategy \cite{OFDM_BX}, 
the equivalent SISO channel, $g$ in \eqref{Eq:Effe2}, can be rewritten as
\begin{align}
\vspace{-12pt}
g&=  (\boldsymbol{\theta}_2 \otimes \boldsymbol{1}_{1\times N_0})^{H} \underline{\boldsymbol{H}}   (\boldsymbol{\theta}_1 \otimes \boldsymbol{1}_{1\times N_0})
\triangleq \boldsymbol{\theta}_2^H \boldsymbol{H} \boldsymbol{\theta}_1,
\vspace{-10pt}
\end{align}
where $\otimes$ denotes the Kronecker product, $\boldsymbol{\theta}_k\!\in\!\mathbb{C}^{M_k\times 1}$ denotes the group-wise reflection vector of IRS $k$, and $\boldsymbol{H}\in\mathbb{C}^{M_2\times M_1}$ denotes the group-wise double-reflection channel from the user to AP with its entries given by $[\boldsymbol{H}]_{j,i} \triangleq \sum_{\ell= (i-1)N_0+1}^{iN_0}\sum_{\tilde{\ell}= (j-1)N_0+1}^{jN_0} [\underline{\boldsymbol{H}}]_{\tilde{\ell},\ell}$, $\forall i\in\mathcal{M}_1\triangleq\{1,\cdots,M_1\}, j\in\mathcal{M}_2\triangleq\{1,\cdots,M_2\}$.
{\bf Transmission protocol:}
We consider a practical transmission protocol, where each channel coherence block of $T$ symbols are divided into two phases. During the first channel-training phase, the user consecutively sends $T_{\rm t}$ pilot symbols to the AP, while the two IRSs properly set their training reflections over time to facilitate the channel estimation for $\boldsymbol{H}$ at  the AP.
Based on the estimated channel  $\widehat{\boldsymbol{H}}$,
the AP first designs the cooperative passive beamforming of the two IRSs for data transmission, denoted by ${\boldsymbol \phi}_1$ and ${\boldsymbol \phi}_2$, respectively, and then feeds them back to the corresponding IRS controllers. After tuning the two IRSs' reflections, 
  the user transmits data over the remaining $T-T_{\rm t}$ symbols in the second phase (with the feedback delay ignored for simplicity). As such, the  average achievable rate in bits/second/Hertz (bps/Hz) per block
   is given by
\begin{align}
\vspace{-20pt}
R&=\!\frac{T-T_{\rm t}}{T}\log_2\l(1\!+\!\frac{|{\boldsymbol \phi}_2^{H}{\boldsymbol H}{\boldsymbol \phi}_1|^2}{\Gamma\sigma^2 }\r),\label{Eq:AchRate}
\vspace{-13pt}
\end{align}
where $\sigma^2\!\triangleq\! \sigma_0^2/P$ denotes the normalized received noise power at the AP with $\sigma_0^2$ and $P$ denoting the noise power and user's transmit power, respectively, and $\Gamma\ge 1$ denotes the achievable rate gap due to a practical modulation and coding scheme. 
\vspace{-5pt}
\section{Proposed Designs for Arbitrary Inter-IRS Channel}\label{Sec:Arbitrary}


In this section, we first propose a general channel estimation scheme (referred to as Scheme $1$) for the double-IRS aided communication system under any arbitrary  inter-IRS channel. Then, the cooperative  passive beamforming of the two IRSs for data transmission is optimized based on the estimated channel.

For the double-IRS channel estimation under any arbitrary inter-IRS channel,  it is worth noting from  \eqref{Eq:Effe2}  that at least $M_1 M_2$ training symbols are required to estimate a total number of $M_1 M_2$  coefficients in the cascaded channel $\boldsymbol{H}$. In the following, we propose Scheme $1$ to estimate $\boldsymbol{H}$ with the minimum training time of $T_{{\rm t},\rm s_1}\triangleq M_1M_2$ symbols. Specifically, the channel training of Scheme $1$ consists of $M_1$ sub-blocks.
In sub-block $1$, we fix the training reflection vector of IRS $1$ as ${\boldsymbol \theta}_1[1]$ and vary the training reflection of  IRS $2$ over $M_2$ training symbols with its training reflection matrix denoted by ${\boldsymbol \Theta}_2\triangleq [{\boldsymbol \theta}_2[1], \cdots,{\boldsymbol \theta}_2[M_2]]\!\in\!\mathbb{C}^{M_2\times M_2}$.
Subsequently, we vary the training reflection of IRS $1$
over the remaining $M_1-1$ sub-blocks;  while in each sub-block, we employ the same (common)  training reflection matrix  (i.e., $\boldsymbol{\Theta}_2$) for IRS $2$. Let $x_{\rm t}$ denote the pilot symbol sent by the user, which is simply set as $x_{\rm t}=1$ without loss of optimality. As such, the received signals at the AP during the channel training with $M_1M_2$ symbols can be stacked into the following matrix form:
\begin{align}
\vspace{-7pt}
 \!\!\!\boldsymbol{Y}_{\rm t}&={\boldsymbol \Theta}_2^H \boldsymbol{H}[{\boldsymbol \theta}_1[1], \cdots, {\boldsymbol \theta}_1[M_1]]  +{\boldsymbol Z}_{\rm t} \triangleq{\boldsymbol \Theta}_2^H \boldsymbol{H}{\boldsymbol \Theta}_1  +{\boldsymbol Z}_{\rm t},\!\!\label{Eq:y2}
 \vspace{-7pt}
 \end{align}
 where $\boldsymbol{Y}_{\rm t}\in\mathbb{C}^{M_2\times M_1}$, ${\boldsymbol \Theta}_1\triangleq [{\boldsymbol \theta}_1[1], \cdots, {\boldsymbol \theta}_1[M_1]]\!\in\!\mathbb{C}^{M_1\times M_1}$, and ${\boldsymbol Z}_{\rm t}\in\mathbb{C}^{M_2\times M_1}$  denotes the received additive white Gaussian noise (AWGN) matrix with  independent and identically distributed (i.i.d.) entries of zero mean and variance $\sigma^2$.

According to \eqref{Eq:y2}, if both ${\boldsymbol \Theta}_1$ and ${\boldsymbol \Theta}_2$ are of full rank, the least-square (LS) estimate of $\boldsymbol{H}$ is given by
\begin{align}
\vspace{-7pt}
\widehat{\boldsymbol{H}}&=({\boldsymbol \Theta}_2^H)^{-1} \boldsymbol{Y}_{\rm t}{\boldsymbol \Theta}_1^{-1}=\boldsymbol{H}+\boldsymbol{H}_{\rm e},\label{Eq:EstH}
\vspace{-7pt}
\end{align}
where $\boldsymbol{H}_{\rm e}\triangleq ({\boldsymbol \Theta}_2^H)^{-1} \boldsymbol{Z}_{\rm t}{\boldsymbol \Theta}_1^{-1}$.
 Let 
$ {\boldsymbol h}_{\rm e}={\rm vec}{({\boldsymbol H}_{\rm e})}
 =({\boldsymbol \Theta}_1^{T}\otimes {\boldsymbol \Theta}_2^H)^{-1} {\rm vec}({\boldsymbol Z}_{\rm t})$.
 The mean square error (MSE) for estimating ${\boldsymbol H}$ in \eqref{Eq:EstH} by Scheme $1$ is given by
 \begin{align}
 \vspace{-7pt}
{\rm MSE}_{{\rm s}_1}&=\mathbb{E}[||{\widehat{\boldsymbol H}}-{\boldsymbol H}  ||_{F}^2]=\mathbb{E}[||{\boldsymbol H}_{\rm e}  ||_{F}^2]=\mathbb{E}[||{\boldsymbol h}_{\rm e} ||^2]\nn\\
&\!\!\overset{(a_1)}{=}\sigma^2 \tr\{({\boldsymbol \Theta}_1^{T}\otimes {\boldsymbol \Theta}_2^H)^{-1} ({\boldsymbol \Theta}_1^{\dag}\otimes {\boldsymbol \Theta}_2)^{-1}\}\nn\\
&\!\!\overset{(a_2)}{=}\sigma^2 \tr\{ ({\boldsymbol \Theta}_1^{\dag}{\boldsymbol \Theta}_1^T )^{-1}\otimes ({\boldsymbol \Theta}_2{\boldsymbol \Theta}_2^H )^{-1} \}\nn\\
&\!\!\overset{(a_3)}{=}\sigma^2 \tr\{ ({\boldsymbol \Theta}_1^{\dag}{\boldsymbol \Theta}_1^T )^{-1}\} \tr\{({\boldsymbol \Theta}_2 {\boldsymbol \Theta}_2^H)^{-1} \},\label{Eq:MSE1}
\vspace{-7pt}
\end{align}
where $(a_1)$ holds since the received noises over different symbols are i.i.d.  and thus $\mathbb{E}[{\rm vec}({\boldsymbol Z}_{\rm t}){\rm vec}({\boldsymbol Z}^H_{\rm t})]=\sigma^2 {\boldsymbol I}_{M_1M_2}$;  $(a_2)$ follows from the matrix operation of $({\boldsymbol A}_1\otimes {\boldsymbol A}_2 ) ({\boldsymbol A}_3\otimes {\boldsymbol A}_4)= ({\boldsymbol A}_1 {\boldsymbol A}_3 )\otimes ({\boldsymbol A}_2 {\boldsymbol A}_4 )$; and $(a_3)$ follows from the matrix operation of $\tr\{{\boldsymbol A}_1\otimes {\boldsymbol A}_2\}=\tr\{{\boldsymbol A}_1\}\tr\{{\boldsymbol A}_2\}$.
Using \eqref{Eq:MSE1},
it can be shown that the 
optimal training reflection matrices for IRSs $1$ and $2$ under the full-rank and unit-modulus constraints should satisfy ${\boldsymbol \Theta}_k{\boldsymbol \Theta}_k^{H}=M_k{\boldsymbol{I}_{M_k}}$ for $k\in\{1,2\}$,
  e.g., 
%
${\boldsymbol \Theta}_k={\boldsymbol D}_{M_k}$,
 where ${\boldsymbol D}_m$ denotes an $m\times m$ discrete Fourier transform (DFT) matrix with its entries given by $\l[{\boldsymbol{D}}_{m}\r]_{\ell,\tilde{\ell}}=e^{-\jmath\frac{2\pi (\ell-1) (\tilde{\ell}-1)}{m}}, \forall 1\le \ell, \tilde{\ell}\le m$. As a result, the minimum MSE of Scheme $1$ is given by ${\rm MSE}_{{\rm s_1},\min}=\sigma^2$.

As for the cooperative passive beamforming design for data transmission, we first observe from \eqref{Eq:AchRate} that maximizing the average achievable rate is equivalent to maximizing the (equivalent SISO) channel power gain $Q=|{\boldsymbol \phi}_2^{H} {\boldsymbol{H}}{\boldsymbol \phi}_1|^2$.
However, since the AP only has imperfect CSI on the cascaded channel, $\widehat{\boldsymbol{H}}$ in \eqref{Eq:EstH}, we instead set the objective to maximize the \emph{expected} (equivalent SISO) channel power gain conditioned on $\widehat{\boldsymbol{H}}$, i.e.,  $\widehat{Q}\triangleq\mathbb{E}_{\boldsymbol{H}_{\rm e}}[Q|\widehat{\boldsymbol{H}}]$. With $\widehat{\boldsymbol{H}}=\boldsymbol{H}+\boldsymbol{H}_{\rm e}$, we have
\begin{align}
\vspace{-15pt}
\widehat{Q}&=\mathbb{E}_{\boldsymbol{H}_{\rm e}}[|{\boldsymbol \phi}_2^{H} {\boldsymbol{H}}{\boldsymbol \phi}_1|^2\big|\widehat{\boldsymbol{H}}]\nn\\
&=\mathbb{E}_{\boldsymbol{H}_{\rm e}}[|{\boldsymbol \phi}_2^{H} \widehat{\boldsymbol{H}}{\boldsymbol \phi}_1-{\boldsymbol \phi}_2^{H} {\boldsymbol{H}_{\rm e}}{\boldsymbol \phi}_1|^2\big|\widehat{\boldsymbol{H}}]\nn\\
&\!\!\overset{(b_1)}{=}|{\boldsymbol \phi}_2^{H} \widehat{\boldsymbol{H}}{\boldsymbol \phi}_1|^2+{\mathbb{E}}_{\boldsymbol{H}_{\rm e}}\l[|{\boldsymbol \phi}_2^{H}{\boldsymbol{H}_{\rm e}}{\boldsymbol \phi}_1|^2\r]\nn\\
&\!\!\overset{(b_2)}{=}|{\boldsymbol \phi}_2^{H} \widehat{\boldsymbol{H}}{\boldsymbol \phi}_1|^2+\sigma^2,\label{Eq:S1AvePower}
\vspace{-15pt}
\end{align}
where $(b_1)$ is due to the independency between $\widehat{\boldsymbol{H}}$ and $\boldsymbol{H}_{\rm e}$, and  $(b_2)$ follows from the definition of $\boldsymbol{H}_{\rm e}$ in \eqref{Eq:EstH} that leads to $\mathbb{E}_{{\boldsymbol{H}_{\rm e}} }[{\boldsymbol{H}_{\rm e}} \boldsymbol{H}_{\rm e}^{H}]=\frac{1}{M_1}({\boldsymbol \Theta}_2^H)^{-1} \mathbb{E}_{{\boldsymbol Z}_{\rm t}}\l[{\boldsymbol Z}_{\rm t}{\boldsymbol Z}_{\rm t}^H\r] {\boldsymbol \Theta}_2^{-1}   =\frac{\sigma^2}{M_2}\boldsymbol{I}_{M_2}$ and ${\mathbb{E}}_{\boldsymbol{H}_{\rm e}}\l[|{\boldsymbol \phi}_2^{H}{\boldsymbol{H}_{\rm e}}{\boldsymbol \phi}_1|^2\r]={\boldsymbol \phi}_2^{H}{\mathbb{E}}_{\boldsymbol{H}_{\rm e}}[  {\boldsymbol{H}_{\rm e}}  {\boldsymbol \phi}_1  {\boldsymbol \phi}_1^H  \boldsymbol{H}_{\rm e}^{H} ]     {\boldsymbol \phi}_2= \frac{\sigma^2}{M_2}{\boldsymbol \phi}_2^{H}\boldsymbol{I}_{M_2}{\boldsymbol \phi}_2=\sigma^2$.  Thus, it can be inferred from \eqref{Eq:S1AvePower} that maximizing $\widehat{Q}$ under the unit-modulus constraints for ${\boldsymbol \phi}_1$ and ${\boldsymbol \phi}_2$ is equivalent to maximizing the channel power gain with respect to (w.r.t.) the estimated channel  $\widehat{\boldsymbol{H}}$, i.e., $|{\boldsymbol \phi}_2^{H} \widehat{\boldsymbol{H}}{\boldsymbol \phi}_1|^2$, since the expected received power associated with the channel estimation error is a constant. 
This optimization problem can be easily shown to be non-convex due to the unit-modulus constraints. To tackle this difficulty, we apply the alternating optimization (AO) method to obtain its suboptimal solution efficiently. Specifically, inspired by the strongest eigenmode beamforming for rate maximization in traditional multiple-input-multiple-output (MIMO) systems,
we first initialize the cooperative passive beamforming for the two IRSs as
 \begin{align}
  \vspace{-8pt}
 [\boldsymbol{\phi}_{1}^{{(0)}}]_{i} =\frac{[\widehat{\boldsymbol{f}}]_{i}}{|[\widehat{\boldsymbol{f}}]_{i}|}, \forall i\in\mathcal{M}_1,~~~
 [\boldsymbol{\phi}_{2}^{{(0)}}]_{j} =\frac{[\boldsymbol{\widehat{d}}]_{j}}{|[\boldsymbol{\widehat{d}}]_{j}|},   \forall j\in\mathcal{M}_2,\nn
 \vspace{-10pt}
 \end{align}
 where $\widehat{\boldsymbol{d}}\in\mathbb{C}^{M_2\times 1}$ and $\widehat{\boldsymbol{f}}^{H}\in\mathbb{C}^{1\times M_1}$ denote the strongest left and right singular vectors of $\widehat{\boldsymbol{H}}$, respectively.  Then, we alternately optimize one of $\{\boldsymbol{\phi}_1, \boldsymbol{\phi}_2\}$  with the other being fixed by using the semidefinite relaxation (SDR) and Gaussian randomization techniques \cite{wu2019intelligent,you2019progressive}; the details are omitted for brevity. It can be easily shown that the proposed AO-based algorithm is guaranteed to converge to at least a locally optimal solution.

\vspace{-8pt}
\section{Proposed Designs for LoS-dominant \\ Inter-IRS Channel}

In this section, we consider the typical scenario with an LoS inter-IRS channel (or LoS-dominant channel in practice).
 In this case, the element-wise inter-IRS channel can be approximated as $\underline{\boldsymbol{G}}_{\rm I}\approx \underline{\boldsymbol{G}}_{\rm I}^{\rm L}=s\underline{\boldsymbol{q}}_2 \underline{\boldsymbol{q}}_1^H,
$
 where $s$ denotes the complex-valued path gain, and $\underline{\boldsymbol{q}}_1^{H}$ and  $\underline{\boldsymbol{q}}_2$ represent the element-wise transmit and receive array response vectors at IRSs $1$ and $2$, respectively \cite{han2020cooperative,you2020fast}.
 As such, the equivalent SISO channel from the user to AP in \eqref{Eq:Effe2} can be rewritten as 
 \begin{align}
 g&=\underline{{\boldsymbol \theta}}_2^{H} \diag(\underline{\boldsymbol{g}}_{\rm A}) s\underline{\boldsymbol{q}}_2 \underline{\boldsymbol{q}}_1^H  \diag(\underline{\boldsymbol{g}}_{\rm U}) \underline{{\boldsymbol \theta}}_1\nn\\
 &\triangleq \underline{{\boldsymbol \theta}}_2^{H} {\underline{\boldsymbol v}}_2 {\underline{\boldsymbol v}}_1^H\underline{{\boldsymbol \theta}}_1= {\boldsymbol \theta}_2^{H}  {\boldsymbol v}_2 {\boldsymbol v}_1^H {\boldsymbol \theta}_1\triangleq {\boldsymbol \theta}_2^{H} \boldsymbol{H}_{\rm L}{\boldsymbol \theta}_1,\label{Eq:LoSchan}
 \end{align} 
 where $\underline{{\boldsymbol v}}_1^H\!\triangleq\!\underline{\boldsymbol{q}}_1^H  \diag(\underline{\boldsymbol{g}}_{\rm U})$, ${\underline{\boldsymbol v}}_2\!\triangleq\!\diag(\underline{\boldsymbol{g}}_{\rm A}) s \underline{\boldsymbol{q}}_2$, $\{{\boldsymbol \theta}_2^{H},  {\boldsymbol v}_2, {\boldsymbol v}_1^H, {\boldsymbol \theta}_1\}$ are the group-wise versions of $\{\underline{{\boldsymbol \theta}}_2^{H},  \underline{{\boldsymbol v}}_2, \underline{{\boldsymbol v}}_1^H, \underline{{\boldsymbol \theta}}_1\}$ \cite{you2019progressive}, and  $\boldsymbol{H}_{\rm L}\triangleq {\boldsymbol v}_2 {\boldsymbol v}_1^H\in\mathbb{C}^{M_2\times M_1}$.
%
It is worth noting that under the LoS inter-IRS channel model, we only need to estimate the two signature channel vectors, i.e., ${\boldsymbol v}_1^H\in\mathbb{C}^{1\times M_1}$ and ${\boldsymbol v}_2\in\mathbb{C}^{M_2\times 1}$, with totally $M_1+M_2$ channel coefficients, which is much smaller than that of the full channel matrix,  $\boldsymbol{H}_{\rm L}$,  with totally $M_1M_2$ channel coefficients for $M_1\ge1,M_2\ge1$. Inspired by this, we propose a new customized  channel estimation scheme (referred to as Scheme $2$) in the following to estimate $\boldsymbol{H}_{\rm L}$, while the effects of NLoS channel components on the channel estimation and passive beamforming performance will be evaluated by simulations in Section~\ref{Sec:Num}.

Specifically, the channel training of Scheme $2$ consists of the following two sub-blocks. 
\begin{itemize}
\item[1)] Sub-block $1$: In this block, we fix the training reflection vector of IRS  $1$ as $\boldsymbol{\theta}_1=\boldsymbol{1}_{M_1\times 1}$, which reduces  the equivalent user-AP SISO channel to $g={\boldsymbol \theta}_2^H {\boldsymbol v}_2 {\boldsymbol v}_1^H \boldsymbol{1}_{M_1\times 1}
= {\boldsymbol \theta}_2^H {\boldsymbol v}_2 V_1^{\dagger} \triangleq {\boldsymbol \theta}_2^H {{\boldsymbol u}}_2$,
where 
\vspace{-5pt}
\begin{align}
V_{1}^{\dagger}\triangleq\sum_{i=1}^{M_1}[\boldsymbol{v}_1^H]_i, ~~{\boldsymbol u}_2\triangleq{\boldsymbol v}_2 V_{1}^{\dagger} \in\mathbb{C}^{M_2\times 1 },\label{Eq:u2}
\vspace{-13pt}
\end{align} 

while a sequence of $M_2$ training reflection vectors of IRS~$2$, denoted by ${\boldsymbol \Theta}_2=[{\boldsymbol \theta}_2[1], \cdots, {\boldsymbol \theta}_2[M_2]]$, are adopted to estimate ${\boldsymbol u}_2$.  
As such, the received signals at the AP over the $M_2$ training symbols of sub-block $1$ can be stacked as
  $\boldsymbol{y}_{{\rm t}}^{(1)}={\boldsymbol \Theta}_2^H {\boldsymbol u}_2 +{\boldsymbol z}_{{\rm t}}^{(1)},$
where $\boldsymbol{y}_{{\rm t}}^{(1)}\in\mathbb{C}^{M_2\times 1}$, and ${\boldsymbol z}_{{\rm t}}^{(1)}\sim\mathcal{N}_c(0, \sigma^2 \boldsymbol{I}_{M_2})$ denotes the AWGN vector at the AP.
Then, the LS estimate of ${\boldsymbol u}_2$ is given by $\widehat{{\boldsymbol u}}_2= ({\boldsymbol \Theta}_2^H)^{-1}  {\boldsymbol y}_{{\rm t}}^{(1)}={\boldsymbol u}_2+{\boldsymbol u}_{2,{\rm e}}$,
where ${\boldsymbol u}_{2,{\rm e}}\triangleq({\boldsymbol \Theta}_2^H)^{-1}{\boldsymbol z}_{{\rm t}}^{(1)}$.
\item[2)] Sub-block $2$:  Following the similar procedures in sub-block $1$, we fix the training reflection vector of IRS $2$ as $\boldsymbol{\theta}_2=\boldsymbol{1}_{M_2\times 1}$ that leads to
  $g=V_{2} {\boldsymbol v}_1^H {\boldsymbol \theta}_1 \triangleq {\boldsymbol u}_1^H {\boldsymbol \theta}_1$,
where 
\vspace{-5pt}
\begin{align}
V_{2}\triangleq\sum_{j=1}^{M_2}[\boldsymbol{v}_2]_j, ~~{\boldsymbol u}_1^H \triangleq V_{2} {\boldsymbol v}_1^H\in\mathbb{C}^{1  \times M_1},\label{Eq:u1}
\vspace{-20pt}
\end{align} 
while we adopt a minimum number of $M_1$ training reflection vectors ${\boldsymbol \Theta}_1=[{\boldsymbol \theta}_1[1], \cdots, {\boldsymbol \theta}_1[M_1]]$ for IRS~$1$  to estimate ${\boldsymbol u}_1^H$. The received signals are stacked as 
$(\boldsymbol{y}^{(2)}_{{\rm t}})^{T}=  {\boldsymbol u}_1^H {\boldsymbol \Theta}_1 +({\boldsymbol z}_{{\rm t}}^{(2)})^T$,
where $\boldsymbol{y}^{(2)}_{{\rm t}}\in\mathbb{C}^{M_1\times 1 }$, and 
  ${\boldsymbol z}_{{\rm t}}^{(2)}\sim\mathcal{N}_c(0, \sigma^2 \boldsymbol{I}_{M_1})$. As such, the LS estimate of ${\boldsymbol u}_1^H$ is given by $\widehat{{\boldsymbol u}}_1^{H}= (\boldsymbol{y}^{(2)}_{{\rm t}})^{T}  {\boldsymbol \Theta}_1^{-1} ={\boldsymbol u}_1^{H}+{\boldsymbol u}^{H}_{1,{\rm e}}$,
where ${\boldsymbol u}^{H}_{1,{\rm e}}\triangleq ({\boldsymbol z}_{{\rm t}}^{(2)})^T {\boldsymbol \Theta}_1^{-1} $.
\end{itemize}
 Using the definitions of ${{\boldsymbol u}}_2$  and ${{\boldsymbol u}}_1^H$ in \eqref{Eq:u2} and \eqref{Eq:u1} , respectively, 
 the double-reflection cascaded 
   channel, $\boldsymbol{H}_{\rm L}$, can be equivalently expressed as
\begin{align}
\vspace{-5pt}
\boldsymbol{H}_{\rm L}={\boldsymbol v}_2 {\boldsymbol v}_1^H={{\boldsymbol u}_2} {\boldsymbol u}_1^H
/{(V_{1}^{\dagger} V_{2})}\triangleq\rho {{\boldsymbol u}_2} {{\boldsymbol u}_1^H},\label{Eq:HL}
 \vspace{-5pt}
\end{align}
where $\rho\!\triangleq1/{(V_{1}^{\dagger} V_{2})}$.
Moreover, it can be inferred from  \eqref{Eq:u2}  and \eqref{Eq:u1} that $U_2\triangleq\sum_{j=1}^{M_2}[{\boldsymbol u}_2]_j=V_{1}^{\dagger} V_{2}$ and 
 $U_1^{\dagger}\triangleq\sum_{i=1}^{M_1}[{\boldsymbol u}_1^H]_i=V_{1}^{\dagger} V_{2}$,  respectively. Based on the above,  
  $\boldsymbol{H}_{\rm L}$ can be estimated as
\begin{align}
\vspace{-7pt}
\widehat{\boldsymbol{H}}_{\rm L}\triangleq&\widehat{{\boldsymbol v}}_2 \widehat{{\boldsymbol v}}_1^H\triangleq{\widehat{\boldsymbol u}_2} \widehat{\boldsymbol u}_1^H/({\widehat{V}_1^{\dagger}\widehat{V}_2})\triangleq\widehat{\rho}{\widehat{\boldsymbol u}_2} \widehat{\boldsymbol u}_1^H,
 \label{Eq:estHL}
 \vspace{-7pt}
\end{align}
where $\widehat{\rho}\triangleq\frac{1}{(\widehat{U}_1^{\dagger}+\widehat{U}_2)/2}$ with $\widehat{U}_1^{\dagger}\triangleq\sum_{i=1}^{M_1}[\widehat{\boldsymbol u}_1^H]_i$ and $\widehat{U}_2\triangleq\sum_{j=1}^{M_2}[\widehat{\boldsymbol u}_2]_j$. Note that we apply $\widehat{V}_{1}^{\dagger} \widehat{V}_{2}=(\widehat{U}_1^{\dagger}+\widehat{U}_2)/2$ to average the estimation noise  for $\rho$ over $\widehat{U}_1^{\dagger}$ and $\widehat{U}_2$.
Based on \eqref{Eq:estHL}, the MSE of Scheme $2$ for estimating $\boldsymbol{H}_{\rm L}$ is given by
\begin{align}
\!\!\!{\rm MSE}_{{\rm s}_2}&\!=\!\mathbb{E}[||{\widehat{\boldsymbol H}}_{\rm L}\!-\!{\boldsymbol H}_{\rm L}  ||_{F}^2]\triangleq\mathbb{E}[|| \underbrace{\widehat{\rho}{\widehat{\boldsymbol u}_2} \widehat{\boldsymbol u}_1^H - \rho{{\boldsymbol u}_2} {\boldsymbol u}_1^H}_{{{\boldsymbol H}_{\rm L, e}}}  ||_{F}^2].\!\!\!
\vspace{-10pt}
\label{Eq:HLe}
\end{align}
Although the MSE in \eqref{Eq:HLe} is highly difficult to characterize due to the coupling $\widehat{\rho}$ and $\{{\widehat{\boldsymbol u}_2}, \widehat{\boldsymbol u}_1^H\}$, we approximate it in the following lemma, with the detailed derivations given  in Appendix~\ref{App:AppMSE2}.

\begin{lemma}\label{Lem:AppMSE2}\emph{The channel estimation error, ${\boldsymbol H}_{\rm L, e}$ in \eqref{Eq:HLe}, can be approximated by
\begin{align}
{\boldsymbol H}_{\rm L, e}\approx  \widehat{\rho}   \l(   \widehat{\boldsymbol u}_2 {\boldsymbol u}^{H}_{1,{\rm e}} + {\boldsymbol u}_{2,{\rm e}} \widehat{\boldsymbol u}_1^H\r).\label{Eq:ErrorAppro1}
\end{align}
 As a result, the MSE of Scheme $2$ in \eqref{Eq:HLe} is approximated by
\begin{align}
\vspace{-5pt}
{\rm MSE}_{{\rm s}_2}&\approx {\rm MSE}_{{\rm s}_2}^{(\rm ap)}\nn\\
&\triangleq \sigma^2 |\widehat{\rho}|^2 \l(||\widehat{\boldsymbol{u}}_2||^2\tr\{ ({\boldsymbol \Theta}_1 {\boldsymbol \Theta}_1^H)^{-1}\}
\right.\nn\\ &\qquad\quad \left.\qquad 
+||\widehat{\boldsymbol{u}}_1^H||^2\tr\{ ({\boldsymbol \Theta}_2 {\boldsymbol \Theta}_2^H)^{-1}\}\r).\nn
\vspace{-5pt}
\end{align}}
\end{lemma}
{\color{black}
}

Using Lemma~\ref{Lem:AppMSE2}, it can be easily shown that the optimal double-IRS training reflection matrices for minimizing ${\rm MSE}_{{\rm s}_2}^{(\rm ap)}$ are given by 
${\boldsymbol \Theta}_k={\boldsymbol D}_{M_k}$ for $k\in\{1,2\}$;  thus the  minimum approximated MSE is given by 
${\rm MSE}^{(\rm ap)}_{{\rm s_2},\min}=\sigma^2 |\widehat{\rho}|^2 \l(||\widehat{\boldsymbol{u}}_2||^2+||\widehat{\boldsymbol{u}}_1^H||^2\r)$,
which is jointly determined by the noise and estimated channels via $\widehat{\boldsymbol{u}}_1^H$, $\widehat{\boldsymbol{u}}_2$, and $\widehat{\rho}$.

With $\widehat{\boldsymbol{H}}_{\rm L}$ and ${\boldsymbol{H}_{\rm L, e}}$, we then introduce an additional  lemma below for designing the cooperative passive beamforming for data transmission, with the derivations given in Appendix~\ref{App:T1T2}.\!

{\color{black}\begin{lemma}\label{Lem:T1T2}\emph{With ${\boldsymbol H}_{\rm L, e}$ approximated in \eqref{Eq:ErrorAppro1} and given fixed passive beamforming vectors $\boldsymbol{\phi}_1$ and $\boldsymbol{\phi}_2$, $\mathbb{E}\l[|{\boldsymbol \phi}_2^{H} {\boldsymbol{H}_{\rm L, e}}{\boldsymbol \phi}_1|^2\r]$ can be approximated by
\begin{align}
\mathbb{E}\l[|{\boldsymbol \phi}_2^{H} {\boldsymbol{H}_{\rm L, e}}{\boldsymbol \phi}_1|^2\r]\approx \sigma^2 |\widehat{\rho}|^2 \l(|{\boldsymbol \phi}_2^H \widehat{\boldsymbol{u}}_2 |^2+ |\widehat{\boldsymbol{u}}_1^H {\boldsymbol \phi}_1|^2\r).\label{App:interf}
\end{align}
}
\end{lemma}}

Using Lemma~\ref{Lem:T1T2}, 
 the expected channel power gain conditioned on the estimated cascaded channel, ${\widehat{\boldsymbol H}}_{\rm L}$, is given by
 \begin{align}
\!\!\!\!\widehat{Q}_{\rm L}&=\mathbb{E}_{\boldsymbol{H}_{\rm L, e}}[|{\boldsymbol \phi}_2^{H} {\boldsymbol{H}}_{\rm L}{\boldsymbol \phi}_1|^2\big|\widehat{\boldsymbol{H}}_{\rm L}]\nn\\
&=|{\boldsymbol \phi}_2^{H} \widehat{\boldsymbol{H}}_{\rm L}{\boldsymbol \phi}_1|^2+{\mathbb{E}}_{\boldsymbol{H}_{\rm L, e}}\l[|{\boldsymbol \phi}_2^{H}{\boldsymbol{H}_{\rm L, e}}{\boldsymbol \phi}_1|^2\r]\nn\\
&\approx {\big| \widehat{\rho} {\boldsymbol \phi}_2^{H} \widehat{\boldsymbol{u}}_2 \widehat{\boldsymbol{u}}_1^H \!{\boldsymbol \phi}_1\big|^2} \!+\! \sigma^2 |\widehat{\rho}|^2 \l(|{\boldsymbol \phi}_2^H \widehat{\boldsymbol{u}}_2 |^2+\! |\widehat{\boldsymbol{u}}_1^H {\boldsymbol \phi}_1|^2\r)\triangleq\widehat{Q}_{\rm L}^{(\rm ap)}\nn\\
&= |\widehat{\rho}|^2\! \l[ |{\boldsymbol \phi}_2^H \widehat{\boldsymbol{u}}_2 |^2 |\widehat{\boldsymbol{u}}_1^H {\boldsymbol \phi}_1|^2 \!+\!  \sigma^2\!\l(|{\boldsymbol \phi}_2^H \widehat{\boldsymbol{u}}_2 |^2\!+\! |\widehat{\boldsymbol{u}}_1^H \!{\boldsymbol \phi}_1|^2\r)\r].\!\!\label{Eq:avePower2}
\vspace{-5pt}
\end{align}
One can observe from \eqref{Eq:avePower2}  that the approximation of the expected channel power gain, $\widehat{Q}_{\rm L}^{(\rm ap)}$, monotonically increases  with both $|{\boldsymbol \phi}_2^H \widehat{\boldsymbol{u}}_2|$ and $|\widehat{\boldsymbol{u}}_1^H {\boldsymbol \phi}_1|$. Thus, it can be easily shown that the optimal cooperative passive beamforming vectors, ${\boldsymbol \phi}_1$ and ${\boldsymbol \phi}_2$, for maximizing $\widehat{Q}_{\rm L}^{(\rm ap)}$ under the unit-modulus constraints are given by
\begin{align}
 [\boldsymbol{\phi}_{1}]_{i} =\frac{[\widehat{\boldsymbol{u}}_1]_{i}}{|[\widehat{\boldsymbol{u}}_1]_{i}|}, i\in\mathcal{M}_1,~~~
 [\boldsymbol{\phi}_{2}]_{j} &=\frac{[\widehat{\boldsymbol{u}}_2]_{j}}{|[\widehat{\boldsymbol{u}}_2]_{j}|},   j\in\mathcal{M}_2,
 \end{align}
 which indicates that the optimal  $\boldsymbol{\phi}_{1}$ and $\boldsymbol{\phi}_{2}$ should align in-phase with the estimated $\widehat{\boldsymbol{u}}_1$ and $\widehat{\boldsymbol{u}}_2$, respectively.

\vspace{-10pt}
\section{Numerical Results}\label{Sec:Num}
\vspace{-1pt}
Numerical results are presented in this section to demonstrate the effectiveness of the proposed channel estimation and cooperative  passive beamforming designs. Under the three-dimensional (3D) Cartesian coordinate system in meter (m), the locations of the user, AP, and centers of IRSs $1$ and $2$ are set as $(1, 20, 0)$, $(1, 0, 0)$,  $(0, 20, 0)$, and $(0,0,0)$, respectively. The azimuth angles of IRSs $1$ and $2$ w.r.t. the $x$-axis are set as $130^{\circ}$ and $30^{\circ}$, respectively. The Rician factors of the user-IRS $1$ and IRS $2$-AP channels, denoted by $K_{{\rm U}}$ and $K_{{\rm A}}$, respectively,  are set as $K_{{\rm U}}=K_{{\rm A}}=20$ dB,  while the inter-IRS channel Rician factor is specified later.
Moreover, the distance-dependent path loss is modeled by $\beta(d)=\beta_0(d/d_0)^{-\alpha}$, where $d$ denotes the individual link distance, $\beta_0=-35$ dB denotes the reference channel power gain at the distance of $d_0=1$~m, and ${\alpha}$ denotes the path loss exponent of the individual link which is set as $\alpha_{\rm U} = \alpha_{\rm A}= 2.2$ for the user-IRS $1$ and IRS $2$-AP links, and $\alpha_{\rm I} = 2.4$ for the inter-IRS link. The two IRSs have the same number of sub-surfaces, set as $M_1=M_2=M$, each consisting of $N_0=10$ reflecting elements.
 Other parameters are set as $\Gamma=9$ dB, $\sigma_0^2=-79$ dBm, and $P=20$ dBm (unless specified otherwise). All simulations results are averaged over $500$ independent Rician fading channel realizations.

\begin{figure}[t]
\centering
\subfigure[NMSE vs. $K_{\rm I}$.]{\label{FigNMSE}
\includegraphics[width=4.175cm]{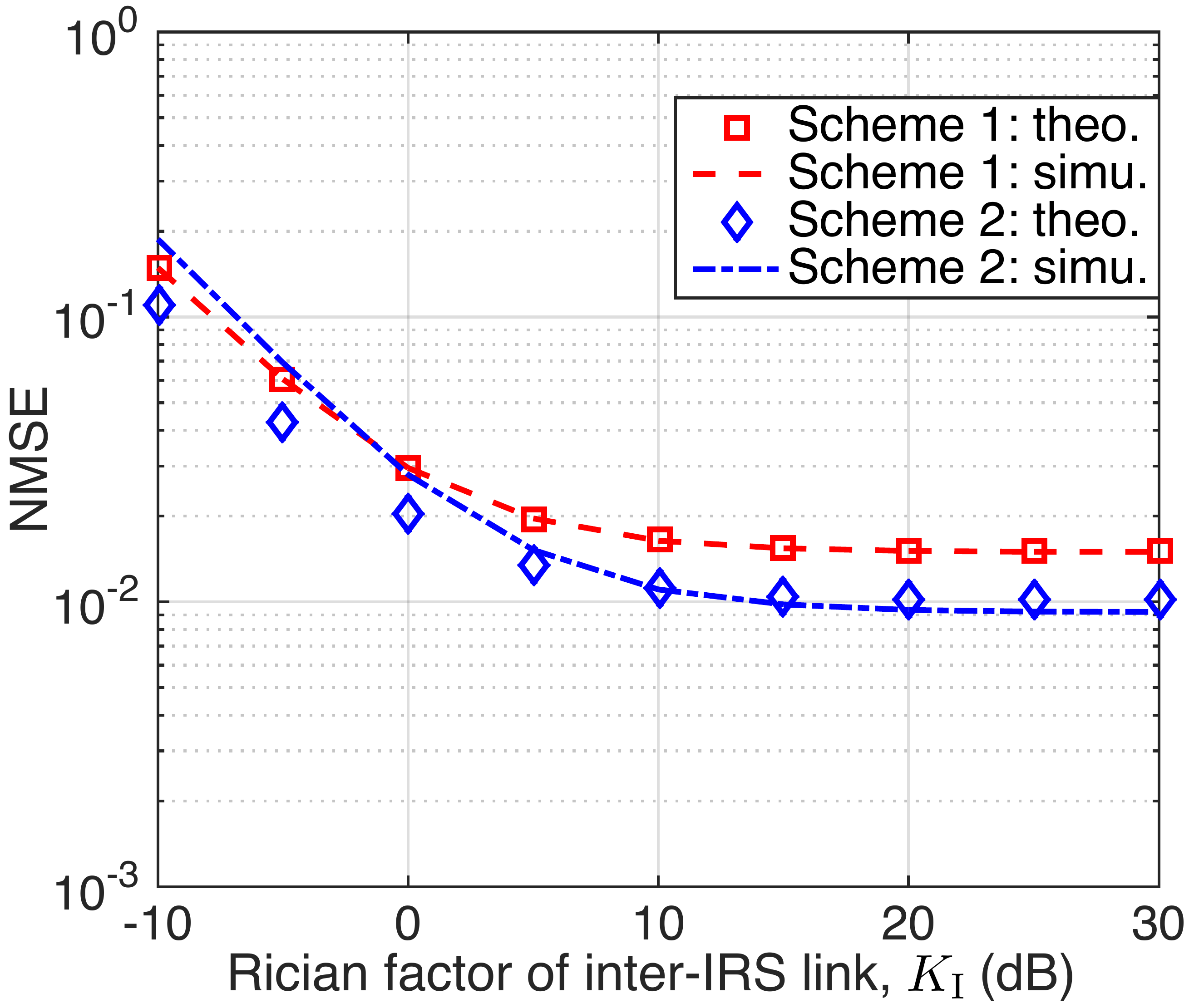}}
\hspace{1.1pt}
\subfigure[Average receive SNR versus $K_{\rm I}$.]{\label{FigSNR}
\includegraphics[width=4.165cm]{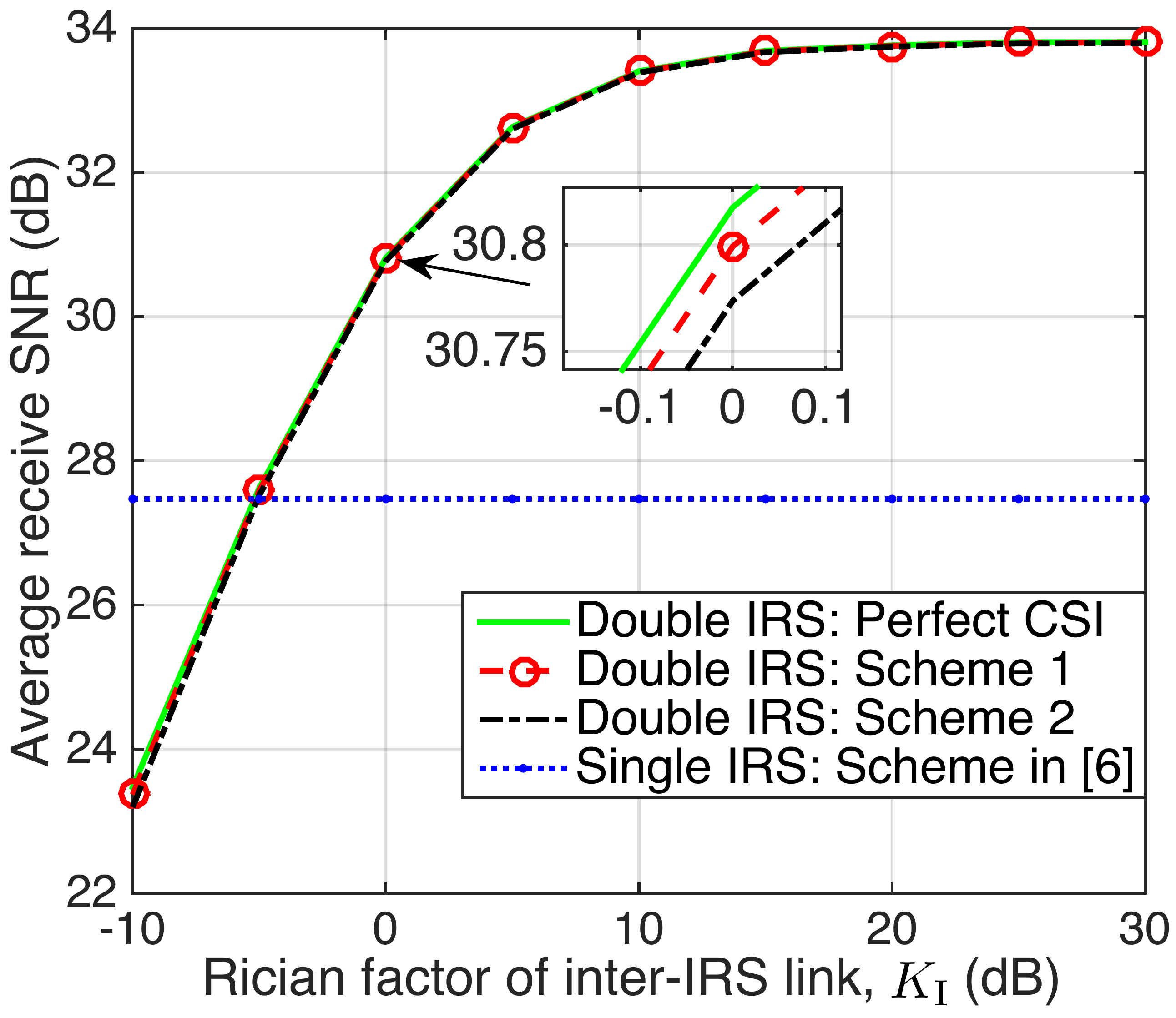}}
\vspace{-5pt}
\caption{Effects of Rician factor $K_{\rm I}$ on the NMSE and average receive SNR with $P=20$ dBm and $M_1=M_2=M=6$.}\label{Fig:2}
\end{figure}

\begin{figure}[t]
\centering
\subfigure[\!Average achievable rate vs. number of sub-surfaces on each IRS, $M$.\!\!\!\!]{\label{FigM}
\includegraphics[width=4.17cm]{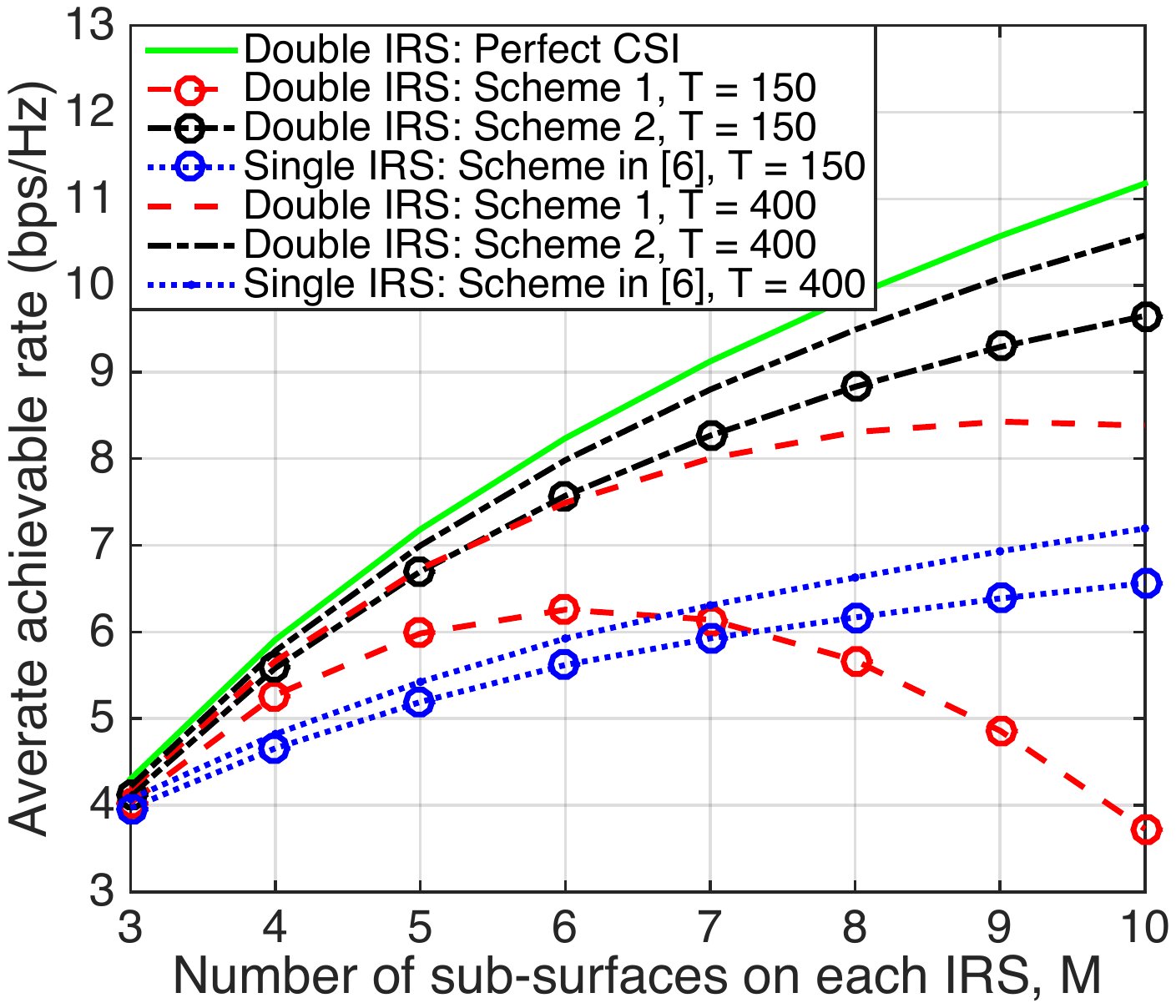}}
\hspace{1pt}
\subfigure[Average achievable rate versus user's transmit power, $P$.]{\label{FigPower}
\includegraphics[width=4.17cm]{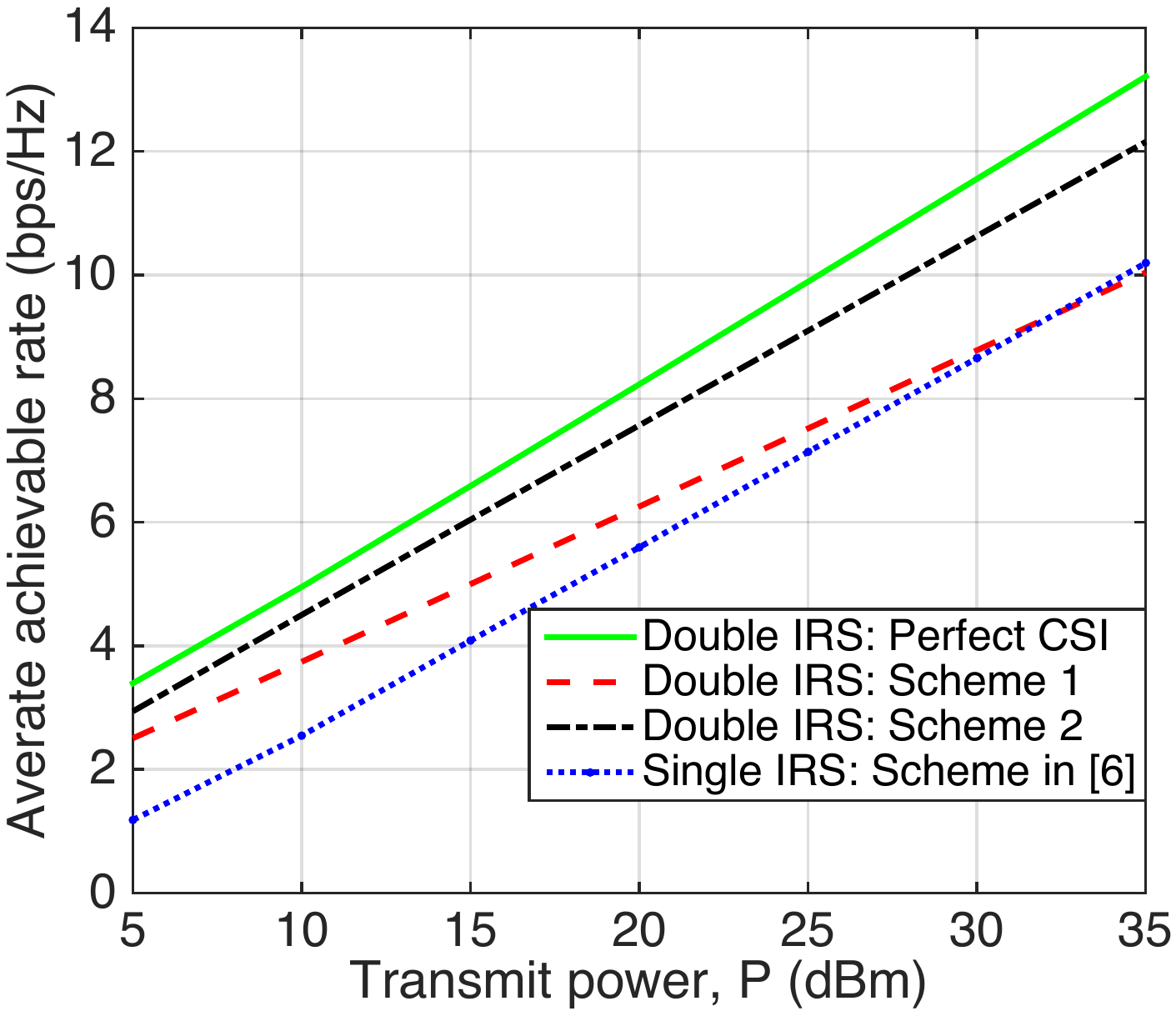}}
\caption{Effects of the number of  sub-surfaces on each IRS, channel coherence time, and user's transmit power with $K_{\rm I}=20$ dB.}
\vspace{-3pt}
\end{figure}

Fig.~\ref{FigNMSE} shows the effects of the inter-IRS channel Rician factor, $K_{\rm I}$, on the normalized MSE (NMSE), which is defined as ${\rm NMSE}=\mathbb{E}[||{\widehat{\boldsymbol H}}-{\boldsymbol H}  ||_{F}^2]/\mathbb{E}[||{\boldsymbol H}  ||_{F}^2]$. It is observed that the theoretical MSE of Scheme $1$ matches well with the simulation results, while that of Scheme $2$ underestimates the MSE in the low-Rician-factor regime as it neglects the effects of NLoS components on the channel estimation error. Moreover, Scheme $2$ requires much less channel training time than Scheme $1$ with $T_{{\rm t},{\rm s_2}}=12$ versus (vs.) $T_{{\rm t},{\rm s_1}}=36$.  In Fig.~\ref{FigSNR}, we compare the average receive signal-to-noise ratio (SNR) at the AP, i.e., $\mathbb{E}[|{\boldsymbol \phi}_2^{H} {\boldsymbol{H}}{\boldsymbol \phi}_1|^2/\sigma^2]$, by the proposed channel estimation and channel gain maximization (CGM)-based passive beamforming designs for the two cooperative IRSs against two benchmarks: 1) double IRS with perfect CSI available at the AP and optimized joint passive beamforming using the proposed AO method; and 2) the conventional single IRS with channel estimation that deploys one traditional IRS by deploying all the  $M_1+M_2=2M$ sub-surfaces at the location of IRS $1$, for which the IRS-AP channel follows the Rayleigh fading with the path loss exponent set as $\alpha_{\rm sin}=4$ (due to more scattering without the double-IRS created LoS path; see Fig.~\ref{Fig:Syst}), while the single-reflection cascaded channel is estimated by the method proposed in \cite{OFDM_BX} with $2M$ training symbols and the passive beamforming is designed based on CGM.
Several interesting observations are made as follows. First, for the double-IRS case, as the Rician factor $K_{\rm I}$ increases, the receive SNR for all schemes first increases and then saturates in the high-Rician-factor regime. Moreover, the average receive SNRs by the  two proposed channel estimation schemes with the optimized passive beamforming 
 both approach to the performance upper bound that assumes perfect CSI, while Scheme $1$ achieves slightly higher receive SNR than Scheme $2$ at the cost of  much longer channel training time (see Fig.~\ref{FigNMSE}). Third, it is observed that deploying two cooperative IRSs significantly outperforms one single IRS when the Rician factor $K_{\rm I}$ is above around $-5$ dB (i.e., even with non-negligible NLoS channel components), since the latter suffers  severe path loss in the  IRS-AP link.

Moreover, we plot in Fig.~\ref{FigM} the average achievable rate vs. the number of sub-surfaces on each IRS, $M$, given different channel coherence time. All schemes employ the CGM-based passive beamforming. It is observed that for the double-IRS case, the proposed channel estimation Scheme $2$ 
achieves much higher rate than Scheme $1$, since they achieve almost the same receive SNR while Scheme $2$ requires much less training time (see Fig.~\ref{Fig:2}).
   In particular, as $M$ increases, the average achievable rate of Scheme $2$ monotonically increases, whereas   that of Scheme $1$ firstly increases and then decreases due to its quadratically-growing channel training time (i.e., $T_{{\rm t},{\rm s_2}}=M^2$). Moreover, given a short channel coherence time (e.g., $T=150$), deploying two cooperative IRSs based on Scheme $2$ is always superior to deploying one single IRS,
and the rate performance gain increases with $M$. Nevertheless, the double-IRS with Scheme $1$ outperforms the single-IRS case only when $M$ is small (e.g., $M<7$), since the channel training time of the former grows much faster than the latter. Furthermore, given a longer channel coherence time (i.e., $T=400$), all schemes achieve enhanced rate performance, among which the double-IRS case with Scheme $1$ attains the maximum rate improvement.

Last, we show  in Fig.~\ref{FigPower} the effects of user's transmit power, $P$, on the average achievable rate for different schemes with $M=6$. It is observed that as $P$ increases, deploying two cooperative IRSs based on channel estimation Scheme $2$ almost achieves a constant rate performance gain as compared to deploying one single IRS. However, the double-IRS case with Scheme $1$ is inferior to the single-IRS case when the transmit power is sufficiently large (e.g., $P=35$ dBm).

\vspace{-8pt}
\section{Conclusions}
\vspace{-2pt}
In this letter, we proposed two different channel estimation schemes for the double-IRS cooperatively aided single-user communication system under different inter-IRS channel setups and optimized their corresponding cooperative passive beamforming for data transmission. By exploiting the rank-one property of the cascaded channel under the LoS inter-IRS channel model, Scheme $2$ was shown to achieve much less channel training time than Scheme $1$, yet without compromising much IRS passive beamforming gain for data transmission, thus attaining a higher average achievable rate. Moreover, simulation results showed that deploying two cooperative IRSs with the proposed channel estimation and cooperative passive beamforming designs significantly outperforms the conventional  single IRS deployed  in the vicinity of the user, especially when the number of IRS reflecting elements is large and/or the inter-IRS channel is LoS-dominant.

\vspace{-3pt}
\appendix
\vspace{-4pt}
\subsection{Proof of Lemma~\ref{Lem:AppMSE2}}\label{App:AppMSE2}
Based on \eqref{Eq:HLe}, we have
\begin{align}
\vspace{-12pt}
\!\!\!\!{\boldsymbol H}_{\rm L, e}&\overset{(c_1)} {\approx}  \widehat{\rho}  \l[ ({\boldsymbol u}_2+{\boldsymbol u}_{2,{\rm e}}) ({\boldsymbol u}_1^H+{\boldsymbol u}^{H}_{1,{\rm e}}) - {{\boldsymbol u}_2} {\boldsymbol u}_1^H\r]\nn\\
&\overset{(c_2)}{\approx} \!\widehat{\rho}  \! \l(   {\boldsymbol u}_2 {\boldsymbol u}^{H}_{1,{\rm e}} + {\boldsymbol u}_{2,{\rm e}} {\boldsymbol u}_1^H\r) \!\approx \! \widehat{\rho}   \l(   \widehat{\boldsymbol u}_2 {\boldsymbol u}^{H}_{1,{\rm e}} + {\boldsymbol u}_{2,{\rm e}} \widehat{\boldsymbol u}_1^H\r),\!\!\label{Eq:ApproHLe}
\vspace{-5pt}
\end{align}
where in $(c_1)$, we assume a small estimation error for $\rho$ such that $\widehat{\rho}\approx \rho$;
 in $(c_2)$, we drop the negligible term $\widehat{\rho} {\boldsymbol u}_{2,{\rm e}} {\boldsymbol u}^{H}_{1,{\rm e}}$.
With \eqref{Eq:ApproHLe}, the MSE of Scheme $2$ in \eqref{Eq:HLe} is approximated by
\vspace{-1pt}
\begin{align}
\vspace{-7pt}
{\rm MSE}_{{\rm s}_2}&
\approx\mathbb{E}\l[ \l\lVert \widehat{\rho} \l(   \widehat{\boldsymbol u}_2 {\boldsymbol u}^{H}_{1,{\rm e}} + {\boldsymbol u}_{2,{\rm e}} \widehat{\boldsymbol u}_1^H\r) \r\rVert_{F}^2 \r]  \label{Eq:ErrorAppro} \\
&\! \overset{(c_3)}{=} |\widehat{\rho}|^2\l(\mathbb{E}\l[\l\lVert  \widehat{\boldsymbol{u}}_2 {\boldsymbol u}^{H}_{1,{\rm e}} \r\rVert_{F}^2  \r]+\mathbb{E}\l[\l\lVert  {\boldsymbol u}_{2,{\rm e}} \widehat{\boldsymbol{u}}_1^H  \r\rVert_{F}^2  \r]\r),
\vspace{-7pt}
\end{align}
where  ($c_3$) holds since  ${\boldsymbol u}_{1,{\rm e}}$ and ${\boldsymbol u}_{2,{\rm e}}$ are independent. By using the definitions of ${\boldsymbol u}_{1,{\rm e}}$ and ${\boldsymbol u}_{2,{\rm e}}$, we can obtain the result in Lemma~\ref{Lem:AppMSE2}.\hfill $\square$
\vspace{-7pt}

\subsection{Proof of Lemma~\ref{Lem:T1T2}}\label{App:T1T2}
\vspace{-3pt}
First, with ${\boldsymbol{H}_{\rm L, e}}\approx  \widehat{\rho}   \l(   \widehat{\boldsymbol u}_2 {\boldsymbol u}^{H}_{1,{\rm e}} + {\boldsymbol u}_{2,{\rm e}} \widehat{\boldsymbol u}_1^H\r)$ in \eqref{Eq:ErrorAppro1},  
$\mathbb{E}\l[  {\boldsymbol{H}_{\rm L, e}}{\boldsymbol \phi}_1{\boldsymbol \phi}_1^H {\boldsymbol{H}^H_{\rm L, e}}  \r]$ can be approximated by
\begin{align}
\vspace{-8pt}
&\mathbb{E}\l[  {\boldsymbol{H}_{\rm L, e}}{\boldsymbol \phi}_1{\boldsymbol \phi}_1^H {\boldsymbol{H}^H_{\rm L, e}}  \r]\nn\\
&\approx |\widehat{\rho}|^2 \l(\mathbb{E}\l[ (\widehat{\boldsymbol{u}}_2 {\boldsymbol u}^{H}_{1,{\rm e}}+{\boldsymbol u}_{2,{\rm e}} \widehat{\boldsymbol{u}}_1^H) {\boldsymbol \phi}_1{\boldsymbol \phi}_1^H (\widehat{\boldsymbol{u}}_2 {\boldsymbol u}^{H}_{1,{\rm e}}+{\boldsymbol u}_{2,{\rm e}} \widehat{\boldsymbol{u}}_1^H)^H  \r]\r)\nn\\
&\!\overset{(d)}{=}|\widehat{\rho}|^2 \l(\mathbb{E}\l[ \widehat{\boldsymbol{u}}_2 {\boldsymbol u}^{H}_{1,{\rm e}}  {\boldsymbol \phi}_1{\boldsymbol \phi}_1^H  {\boldsymbol u}_{1,{\rm e}}  \widehat{\boldsymbol{u}}^H_2\r] +\mathbb{E}\l[ {\boldsymbol u}_{2,{\rm e}} \widehat{\boldsymbol{u}}_1^H {\boldsymbol \phi}_1{\boldsymbol \phi}_1^H  \widehat{\boldsymbol{u}}_1 {\boldsymbol u}^H_{2,{\rm e}} \r]\r)\nn\\
&=\sigma^2 |\widehat{\rho}|^2 \l( \widehat{\boldsymbol{u}}_2 \widehat{\boldsymbol{u}}_2^H + \frac{\widehat{\boldsymbol{u}}_1^H {\boldsymbol \phi}_1{\boldsymbol \phi}_1^H  \widehat{\boldsymbol{u}}_1}{M_2}\boldsymbol{I}_{M_2}\r)\label{Eq:LT2approx},
\vspace{-8pt}
\end{align}
where ($d$) is due to the independency between ${\boldsymbol u}_{1,{\rm e}}$ and ${\boldsymbol u}_{2,{\rm e}}$.

Next, with \eqref{Eq:LT2approx}, we have
\begin{align}
\vspace{-5pt}
&\mathbb{E}\l[|{\boldsymbol \phi}_2^{H} {\boldsymbol{H}_{\rm L, e}}{\boldsymbol \phi}_1|^2\r]\nn\\
\approx~& {\boldsymbol \phi}_2^{H}\l(\sigma^2  |\widehat{\rho}|^2\l( \widehat{\boldsymbol{u}}_2 \widehat{\boldsymbol{u}}_2^H + \frac{\widehat{\boldsymbol{u}}_1^H {\boldsymbol \phi}_1{\boldsymbol \phi}_1^H  \widehat{\boldsymbol{u}}_1}{M_2}\boldsymbol{I}_{M_2}\r)\r) {\boldsymbol \phi}_2\nn\\
=~&\sigma^2 |\widehat{\rho}|^2 \l(|{\boldsymbol \phi}_2^H \widehat{\boldsymbol{u}}_2 |^2+ |\widehat{\boldsymbol{u}}_1^H {\boldsymbol \phi}_1|^2\r),
\end{align}
thus completing the proof.			\hfill $\square$
\vspace{-2pt}

\bibliographystyle{IEEEtran}
\bibliography{BibDesk_File}

\end{document}